\documentclass[twocolumn,showpacs,preprintnumbers,amsmath,amssymb,nofootinbib]{revtex4}
\usepackage{graphics}
\usepackage{graphicx}
\usepackage{epstopdf}
\usepackage{comment}
\usepackage{authordate1-4}
\begin{document}
\title{Mechanisms of pattern formation during T cell adhesion}
\author{Thomas R.\ Weikl\footnote{email: Thomas.Weikl@mpikg-golm.mpg.de} and 
             Reinhard Lipowsky\footnote{email: Reinhard.Lipowsky@mpikg-golm.mpg.de}}
\affiliation{Max-Planck-Institut f\"ur Kolloid- und Grenzfl\"achenforschung, 14424 Potsdam, Germany}

\begin{abstract}
T cells form intriguing patterns during adhesion to antigen-presenting cells. The patterns at the cell-cell contact zone are composed of two types of domains, which either contain short TCR/MHCp receptor-ligand complexes or the longer LFA-1/ICAM-1 complexes. The final pattern consists of a central TCR/MHCp domain surrounded by a ring-shaped LFA-1/ICAM-1 domain, while the characteristic pattern formed at intermediate times is inverted with TCR/MHCp complexes at the periphery of the contact zone and LFA-1/ICAM-1 complexes in the center. In this article, we present a statistical-mechanical model of cell adhesion and propose a novel mechanism for the T cell pattern formation. Our mechanism for the formation of the intermediate inverted pattern is based (i) on the initial nucleation of numerous TCR/MHCp microdomains, and (ii) on the diffusion of free receptors and ligands into the contact zone. Due to this inward diffusion, TCR/MHCp microdomains at the rim of the contact zone grow faster and form an intermediate peripheral ring for sufficiently large TCR/MHCp concentrations.  In agreement with experiments, we find that the formation of the final pattern with a central TCR/MHCp domain requires active cytoskeletal transport processes. Without active transport, the intermediate inverted pattern seems to be metastable in our model, which might explain patterns observed during natural killer (NK) cell adhesion. At smaller TCR/MHCp complex concentrations, we observe a different regime of pattern formation with intermediate multifocal TCR/MHCp patterns  which resemble experimental patterns found during thymozyte adhesion.
\end{abstract}
\maketitle

\section{Introduction}

T lymphozytes mediate immune responses by adhering to cells which display foreign peptide fragments on their surface. These peptide fragments are presented by MHC molecules on the cell surfaces, and recognized by the highly specific T cell receptors (TCR). Cytotoxic T cells directly kill virally infected cells after adhesion,  while helper T cells help to activate other lymphozytes, e.g.\ B cells which produce and secrete antibodies.

At the contact zone of a T cell with an antigen-presenting cell (APC), bound receptor-ligand pairs are arranged in characteristic supramolecular patterns, the `immunological synapse' \cite{monks98,grakoui99,krummel00,potter01}, for reviews see \cite{vanderMerwe00, dustin00,delon00,bromley01,dustin01,wulfing02}.  A current focus in T cell biology is to understand the formation of these patterns and their role for T cell activation \cite{khlee02}. The final pattern of an adhering T cell consists of a central domain with bound TCR/MHCp complexes, surrounded by a ring-shaped domain in which the integrin receptors LFA-1 of the T cell are bound to their ligands ICAM-1 of the APC. Intriguingly, the characteristic intermediate pattern formed earlier during T cell adhesion is inverted, with a TCR/MHCp ring surrounding a central LFA-1/ICAM-1 domain in the contact zone \cite{grakoui99,johnson00,khlee02}. This pattern inversion has been first observed for T cells adhering to a supported lipid bilayer with MHCp and ICAM-1, \cite{grakoui99,johnson00}, more recently also in a cell conjugate system \cite{khlee02}. In the case of natural killer (NK) cells, the inverted synapses seem to be stable \cite{davis99,fassett01}. More recently, thymozytes (immature T cells), have been found to form multifocal synapses with several nearly circular clusters of TCR/MHC-peptide complexes in the contact zone \cite{hailman02,richie02}. 

As has been proposed by several groups, the lateral segregation or phase separation in the immunological synapse probably is caused by the length difference between receptor/ligand complexes \cite{davis96,shaw97,vanderMerwe00,qi01,sjlee02,weikl02a,burroughs02,chen03,coombs}. Bound TCR/MHCp complexes induce a membrane separation of about 15~nm, while LFA-1/ICAM-1 complexes have a larger extension of 40~nm \cite{dustin00}. In general,  the lateral phase behavior of the membranes is also affected by the concentrations of the complexes \cite{weikl02a,weikl01,weikl02b}. Lateral phase separation only occurs if the complex concentrations exceed a critical threshold.  
 An additional driving force for phase separation comes from large glycoproteins such as CD43 and CD45 \cite{weikl02b}. These glycoproteins have a length of 40~nm and more \cite{shaw97} and thus form a steric barrier for TCR/MHCp binding.

To understand the time-dependent pattern evolution, it is necessary to consider the dynamics of the phase separation process in the cell geometry. The first theoretical model based on a Landau-Ginzburg free energy coupled to a set of reaction-diffusion equation has been proposed by Qi, Groves, and Chakraborty \cite{qi01,sjlee02}. Qi et al.\ obtain circularly symmetric adhesion patterns exhibiting the characteristic domain inversion and conclude that the entire T cell pattern evolution may be caused by self-assembly.  The formation of the intermediate inverted synapse is suggested to be formed in a pivot mechanism in which the formation of  LFA-1/ICAM-1 complexes in the center of the cell contact zone causes close apposition in the periphery allowing TCR/MHCp binding.

Here, we consider a statistical-mechanical model for the cell adhesion dynamics in which the T cell and APC membranes are discretized into small patches of linear size $a\simeq 70$ nm. In this model, the configurational energy of the membranes depends on the numbers of receptors, ligands, and glycoproteins present in each patch, and the separation of apposing patches in the cell contact zone. The adhesion dynamics is studied with Monte Carlo simulations, which include the whole range of fluctuations in membrane shape and composition. Based on the patterns observed in the simulation, we propose a novel mechanism for the formation of the intermediate inverted synapse. We observe an initial formation of  many small TCR/MHCp microclusters throughout the contact zone. Subsequently, TCR/MHPp clusters at the periphery of the contact zone grow faster due to the diffusion of free receptors and ligands into this zone. For sufficiently large TCR/MHCp concentrations, these peripheral clusters coalesce into a closed ring, surrounding a central domain of LFA-1/ICAM-1 complexes. At smaller TCR/MHCp concentrations, we observe a different dynamic regime with characteristic multifocal intermediate patterns consisting of several circular TCR/MHPp domains, which resemble patterns observed during thymozyte adhesion. 

Our proposed mechanism for the formation of the intermediate inverted T cell synapse is a self-assembly mechanism based on TCR/MHCp microcluster nucleation and the diffusion of free receptors and ligands into the contact zone. However, we find that the final, `mature' T cell synapse only arises in the presence of  active cytoskeletal transport of TCRs to the center of the contact zone. This seems to be in agreement with experimental findings. Cytochalasin D, an inhibitor of actin-based transport,  inhibits also the central TCR/MHCp movement \cite{grakoui99}. An active actin/myosin-based transport of receptors into the contact zone has been observed \cite{wulfing98a}, while the glycoprotein CD43 is actively moved out of the contact zone \cite{allenspach01,delon01}.
Without active transport, the intermediate TCR/MHCp ring seems to be metastable and persists for an hour and more, according to our simulations. This might explain the inverted synapse of natural killer (NK) cells which consists of a peripheral ring of short HLA-C/KIR complexes surrounding a central domain containing the longer LFA-1/ICAM-1 complexes. The formation of the inverted NK synapse is not inhibited by ATP depletion or disruption of the cytoskeleton and thus appears to be caused by self-assembly \cite{davis99,fassett01}. 

\section{Model}

%
\begin{figure}
\resizebox{\columnwidth}{!}{\includegraphics{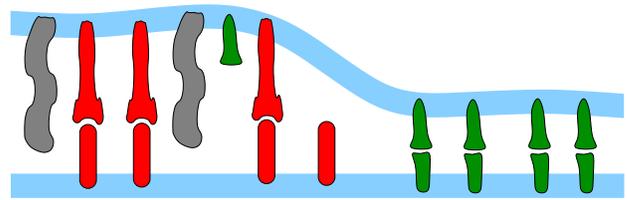}}
\caption{Cartoon of a T cell membrane (top) adhering to an APC membrane (bottom). The T cell membrane contains the T cell receptor TCR (green) and the receptor LFA-1 (red). The APC membrane contains the corresponding ligands MHCp (green) and  ICAM-1 (red). Both membranes contain repulsive glycoproteins (grey). Because of the different lengths of bound TCR/MHCp complexes, LFA-1/ICAM-1 complexes, and glycoproteins, the membrane phase separates into domains. \label{cartoon}} 
\end{figure}

In this section, we describe our theoretical model for the interaction of a T cell with an antigen-presenting cell (APC). We consider two apposing membranes. The first membrane represents the T cell and contains the receptors TCR and LFA-1. The second membrane represents the APC and contains the ligands MHCp and ICAM-1. We use the terms `receptors' and `ligands' here with respect to the T cell: Adhesion molecules anchored in the T cell membrane are called `receptors', and those in the APC membrane are `ligands'. Protruding glycoproteins are embedded in both membranes, forming a steric barrier for the formation of the short TCR/MHCp complexes (see Fig.~\ref{cartoon}). 

\begin{figure}
\resizebox{0.5\columnwidth}{!}{\includegraphics{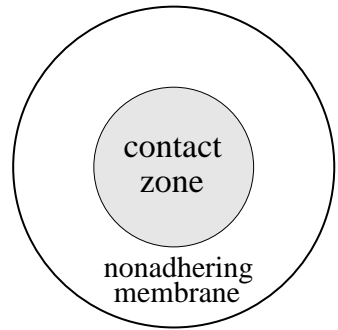}}
\caption{`Cell' adhesion geometry: The circular contact zone is surrounded by a nonadhering membrane ring. Receptors, ligands, and glycoprotein diffuse around in the whole membrane, but interact with the apposing membrane only within the contact zone. \label{geometry}}
\end{figure}

To mimick the adhesion geometry of the cells, we divide the membranes into a contact zone and a surrounding region in which the membranes do not interact. The receptors can diffuse in the whole T cell membrane, but interact with the ligands of the APC membrane only within the contact zone of the two membranes. For simplicity, we avoid the problem of modeling the full cell shape, and assume here that the contact zone has an essentially circular shape and a constant area on the time scales considered here, see Fig.~\ref{geometry}. This contact zone is thought to be established in fast initial adhesion events after first cell contact.  Experimental pictures of adhering T cells show that the contact zone fully develops in less than 30 seconds \cite{grakoui99}.

To characterize the membrane conformations, we partition both membranes into quadratic patches with linear extension $a$.\footnote{More precisely, we are discretizing the reference plane shown in Fig.~\ref{geometry} into a square lattice with lattice constant $a$, which results in partitioning both membranes into quadratic patches.} The local composition of the T cell membrane is then described by the numbers $n_i^T$ of TCRs, $n_i^L$ of LFA-1, and $n_i^{Gt}$ of glycoproteins in each membrane patch $i$. Correspondingly, the composition of the APC membrane is given by the numbers $n_i^M$ of MHCp, $n_i^{I}$ of ICAM-1, and $n_i^{Ga}$ of glycoproteins in all patches. Within the contact zone, the local separation between two apposing membrane patches of the two cells is denoted by $l_i$.

The elastic energy of the membranes in the contact zone is dominated by the bending energy and by lateral tension and can be written as
\begin{equation}
{\cal H_\text{el}}\{l\} = \sum_i\left[(\kappa/2a^2)(\Delta_d l_i)^2 + (\sigma/2)(\nabla_d l_i)^2\right]
\label{elasticEnergy}
\end{equation}
Here, $\kappa=\kappa_1\kappa_2/(\kappa_1+\kappa_2)$ denotes the effective bending rigidity of the two membranes with rigidities $\kappa_1$ and $\kappa_2$, and $\sigma$ is a lateral tension. For simplicity, the effective bending rigidity is taken to be independent of the local membrane composition. The term
\begin{displaymath}
\Delta_d l_i=\Delta_d l_{x,y} =l_{x+a,y}+l_{x-a,y}+l_{x,y+a}+l_{x,y-a}-4l_{x,y}
\end{displaymath}
is the total curvature of the membrane separation field $l_i$ at site $i$, and
\begin{displaymath}
(\nabla_d l_i)^2=(\nabla_d l_{x,y})^2=(l_{x+a,y}-l_{x,y})^2 + (l_{x+a,y}-l_{x,y})^2
\end{displaymath}
describes the local area increase of the curved membranes with respect to the reference $x$-$y$ plane given by $l_i=l_{x,y}=0$. The elastic energy (\ref{elasticEnergy}) dominates the fluctuations  of the membrane separation in the contact zone, whereas the overall cell shape is also affected by the elasticity of the cytoskeleton which is coupled to the membranes. In the simulations, we use the dimensionless separation field $z=(l/a)\sqrt{\kappa/(k_B T)}$, and choose the value $z=1$ to correspond to a length of 20 nm, which results in the relation $a=20\sqrt{\kappa/(k_B T)}$~nm for the linear patch size. For the typical rigidities $\kappa_1=\kappa_2=25 k_B T$ of the two biomembranes, the effective rigidity $\kappa$ has the value $12.5 k_BT$, and the linear patch size is $a\simeq 70$ nm. For the lateral tension, we choose the value $\sigma=0.1 \kappa /a^2\simeq 2\cdot 10^{-6}N/m$.

The overall configurational energy of the membranes in the contact zone is the sum of the elastic energy (\ref{elasticEnergy}) and the interaction energies of receptors, ligands, and glycoproteins:
\begin{eqnarray}
{\cal H}\{l,n\} = {\cal H_\text{el}}\{l\} + \sum_i\Big[V_\text{hw} + \min(n_i^T,n_i^M)V_\text{TM}(l_i)\nonumber\\
  +  \min(n_i^L,n_i^I)V_\text{LI}(l_i) + \left(n_i^{Gt}+n_i^{Ga}\right) V_G(l_i)\Big]
\label{totalEnergy}
\end{eqnarray}
Here, $V_\text{TM}(l_i)$ and $V_\text{LI}(l_i)$ are the attractive interaction potentials of TCR/MHCp and LFA-1/ICAM-1 complexes, respectively, while $V_G(l_i)$ is the repulsive interaction potential of the glycoproteins. The term $\min(n_i^T,n_i^M)$ denotes the minimum of the numbers of TCR and MHCp molecules at site $i$. This minimum is equivalent to the number of interacting TCR/MHCp pairs in the apposing patches at site $i$.

\begin{figure*}
\vspace{-0.2cm}
\resizebox{2\columnwidth}{!}{\includegraphics{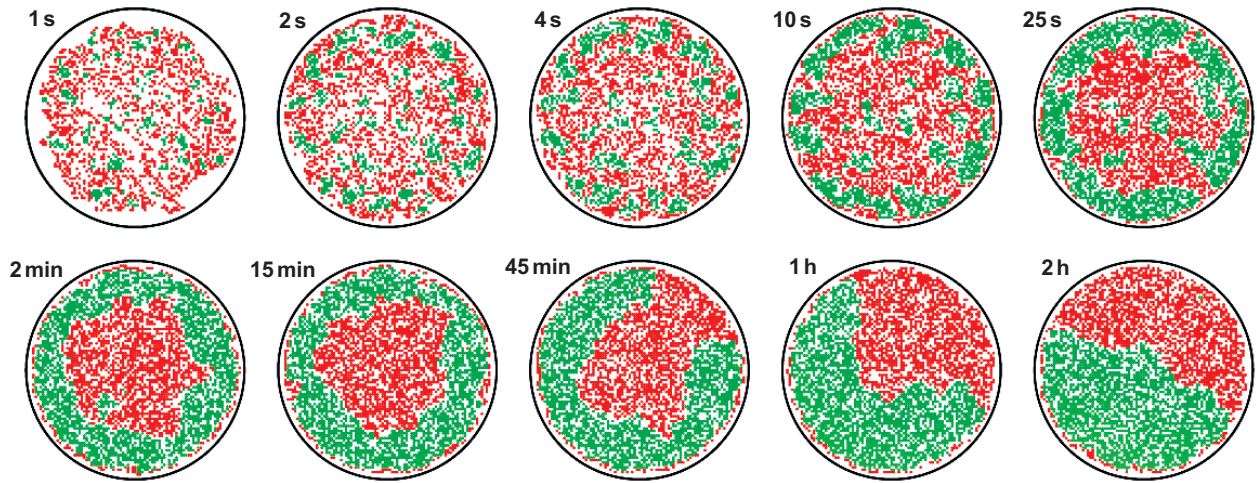}}
\vspace{-1.3cm}
\caption{Time sequence of Monte Carlo conformations of the contact zone for the effective binding energies $U_{TM}=-6.5 k_B T$ of TCR/MHCp complexes and  $U_{LI}=-3 k_B T$ of LFA-1/ICAM-1 complexes. The overall concentrations of TCR, ICAM-1, LFA-1, and glycoproteins in each of the membranes is $0.4 / a^2 \simeq 80$ molecules/$\mu$m$^2$ for a linear patch size $a\simeq 70$ nm, and the concentration of MHCp is $0.1 / a^2 \simeq 20$ molecules/$\mu$m$^2$. Membrane patches with bound TCR/MHCp complexes are shown in green, patches with bound LFA-1/ICAM-1 complexes in red. }
\label{rings}
\end{figure*}

The receptor-complexes can only form if the membrane separation is in an appropriate range. The length of the TCR/MHCp complexes is about 15~nm, while the LFA-1/ICAM-1 complexes have a length of about 40~nm. Since the membrane within a patch is `rough' due to the thermal fluctuations on length scales smaller than the linear extension $a\simeq 70$~nm of the patches, we assume that the complexes can form if the separation of two apposing patches does not deviate more than $5$~nm from the lengths  $z_{TM}$ and $z_{LI}$. The interaction potential of TCR and MHCp is then given by
\begin{eqnarray}
 V_{TM} &=& U_{TM}  \hspace{0.3cm} \text{for} \hspace{0.2cm} 10\text{~nm} < l_i < 20\text{~nm},\nonumber\\
      &=& 0 \hspace{0.9cm} \text{otherwise} \label{potTM}
\end{eqnarray}
 and the interaction potential of ICAM-1 and LFA-1 is
\begin{eqnarray}
 V_{LI} &=& U_{LI}  \hspace{0.3cm} \text{for} \hspace{0.2cm} 35\text{~nm} < l_i < 45\text{~nm},\nonumber\\
      &=& 0 \hspace{0.9cm} \text{otherwise} \label{potLI}      
\end{eqnarray}
where $U_{TM}<0$ is the binding energy of a TCR/MHCp complex, and $U_{LI}<0$ the binding energy of LFA-1/ICAM-1. As noted above, the potential width of 10~nm effectively takes into account small-scale fluctuations within patches. Thus, this width does not result from the atomic interaction potentials of receptor and ligand molecules, which should have a significantly smaller range. Similarly, the binding energies of the receptor/ligand complexes should be seen as effective binding energies which can be used to adjust the 2-dimensional equilibrium constants of the TCR/MHCp and LFA-1/ICAM-1 complexes. The 2-dimensional equilibrium constants are approximately given by $K_{TM}\simeq a^2 e^{-U_{TM}/(k_B T)}$ and $K_{LI}\simeq a^2 e^{-U_{LI}/(k_B T)}$ (see Appendix). 

The repulsive glycoproteins protruding from both membranes vary in size. However, many of these proteins have a length comparable to the length of the LFA-1/ICAM-1 complexes. These glycoproteins do not inhibit the binding of ICAM-1 and LFA-1, but impose a steric barrier for the formation of TCR/MHCp complexes. They are characterized here by the potential
 \begin{eqnarray}
 V_{G} &=& U_G (l-l_G)^2    \hspace{0.3cm} \text{for} \hspace{0.2cm} l < l_G,\nonumber\\
      &=& 0 \hspace{0.9cm} \text{otherwise}
 \end{eqnarray}
with $U_G=10\kappa/a^2$ and $l_G=40$~nm. This potential results from the fact that a membrane patch of size $a$ containing a glycoprotein has to bend around this protein to achieve an overall patch separation smaller than the length of the glycoprotein.

In the following, the radius of the circular contact zone is chosen to be 45$a$, and the nonadhering membrane surrounding the contact zone is a ring of width $55a$. As boundary condition at the rim of the contact zone, the membrane separation is fixed at a value of 100~nm, which is significantly larger than the length of the TCR-MHCp and LFA-1/ICAM-1 complexes and the glycoproteins. 

\section{Adhesion dynamics in the absence of cytoskeletal transport processes}

We first consider the pattern formation in the absence of active forces which transport molecules in or out of the contact zone. The lateral motion of receptors, ligands, and glycoproteins within the membranes is then purely diffusive. In our discretized membranes, the diffusive motion of the macromolecules is modeled as a hopping process between neighboring membrane patches. Each receptor, ligand, or glycoprotein in a certain membrane patch can hop to one of the four nearest neighbor patches during a single time step. The hopping processes of macromolecules located in the nonadhering membrane region do not change the cell interaction energy, eq.~(\ref{totalEnergy}). However, within the contact zone, the attempted hopping of a macromolecule may change the free energy. According to the standard Metropolis criterion, the hopping attempt is always accepted if it does not increase the free energy, but is only accepted with probability $\exp(-\Delta F/(k_B T))$ if it leads to a free energy increase $\Delta F$.\footnote{We reject moves in which bound ligands or receptors hop from one binding partner to another in a single time step. For these moves, the free energy difference would be zero. Thus, the actual free energy barrier for the unbinding process of the ligand/receptor complex would not be captured.} During a time step, we also attempt to shift the separation $l_i$ between apposing membrane patches in the contact zone by $d\cdot\zeta[-1,1]$ where $d$ is the step width $10$~nm, and $\zeta[-1,1]$ is a random number between $-1$ and $1$. 

A single Monte Carlo step roughly corresponds to 1 ms of real time. This time estimate can be derived from the 2-dimensional diffusion law $\langle x^2\rangle = 4 D t$ and the typical diffusion constant $D\simeq 1$~$\mu$m$^2$/s for membrane-anchored macromolecules. In a single Monte Carlo step, a free receptor, free ligand, or a glycoprotein moves a distance $a$ to a neighboring membrane patch, which corresponds to a diffusion time $t=a^2/(4 D)\simeq 1$ ms for $a=70$~nm. On the length scale of our patches, the diffusive motion of the macromolecules is slower than the relaxation of the membrane separation \cite{brochard75} and hence defines the time scale.

\begin{figure*}[t]
\vspace{-0.2cm}
\resizebox{2\columnwidth}{!}{\includegraphics{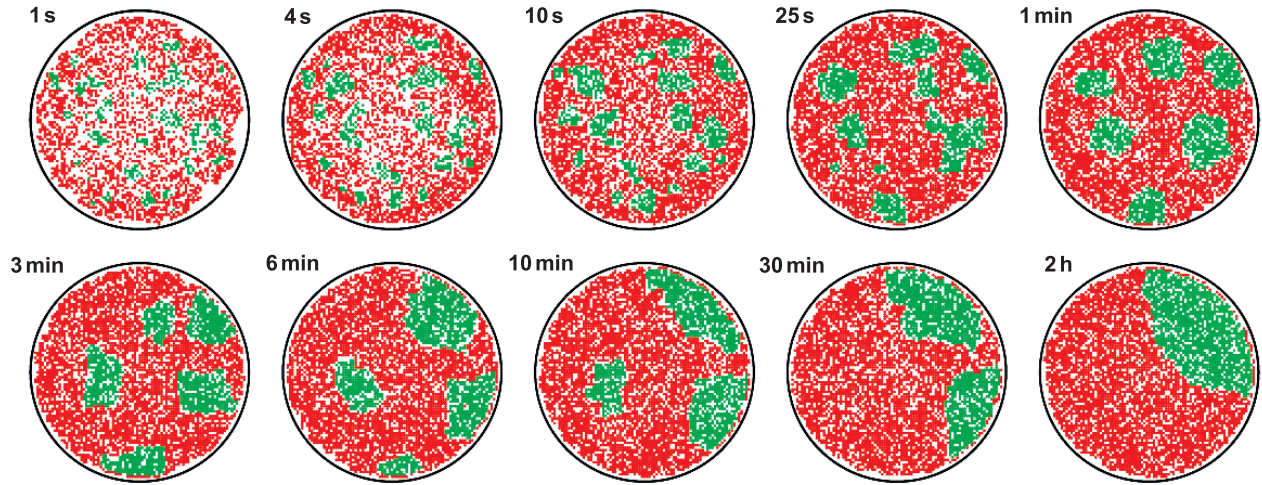}}
\vspace{-1.3cm}
\caption{Time sequence of Monte Carlo conformations of the contact zone for the effective binding energies $U_{TM}=-5.5 k_B T$ of TCR/MHCp complexes,  $U_{LI}=-4 k_B T$ of LFA-1/ICAM-1 complexes, and the same molecular concentrations as in Fig.~\ref{rings}. Membrane patches with bound TCR/MHCp complexes are shown in green, patches with bound LFA-1/ICAM-1 complexes in red.  
}
\label{multifocal}
\end{figure*}

These dynamic rules and the free energy given in eq.~(\ref{totalEnergy}) specify a stochastic adhesion process. Here, we study the adhesion process with Monte Carlo simulations. Taking averages over many independent Monte Carlo runs gives a numerical solution of the corresponding master equation \cite{vanKampen92,binder92}. The stochastic process captures the fluctuations in the membrane separation and describes the diffusive motion of the receptors, ligands, and glycoprotein on a single-molecule level, which is essential for the mechanisms of pattern formation considered in this article.  
As initial conformation, we choose the separation profile $l=l_o +c r^4$ where $r$ is the distance from the center of the contact zone, $l_o$ is 45~nm, and $c>0$ is chosen so that the separation at the rim of the contact zone with radius $r=45a$ is 100~nm (boundary condition). This initial separation in the contact zone is larger than 45~nm, and thus beyond the interaction range of receptors, ligands, and glycoproteins. Initially, these molecules are taken to be randomly distributed within the whole membrane.

We systematically study the adhesion dynamics for various concentrations of the receptors, ligands, and glycoproteins and for various effective binding energies, or 2d equilibrium constants, of the TCR/MHCp and LFA-1/ICAM-1 complexes. Since the length difference of the complexes leads to phase separation at the molecular concentrations considered here, the two types of receptor/ligand complexes have to `compete' for the contact zone. In general, the overall area of TCR/MHCp domains in the contact zone increases with the concentrations of TCR and MHCp molecules and with the effective binding energy $U_{TM}$.  However, if the molecular concentrations or the binding energy are too small, TCR/MHCp domains do not form, and the contact zone contains only bound LFA-1/ICAM-1 complexes. At molecular concentrations and binding energies where TCR/MHCp and LFA-1/ICAM-1 domains coexist, we observe two different regimes for the dynamics with clearly distinct patterns of TCR/MHCp domains at intermediate times. The pattern evolution roughly depends on the overall area of TCR/MHCp domains after initial relaxation.

{\it Regime 1}: If the overall area of TCR/MHCp domains is relatively large, we observe a characteristic ring-shaped TCR/MHCp domain at intermediate times, surrounding a central domain of LFA-1/ICAM-1 complexes. A typical example for the pattern evolution in this regime is presented in Fig.~\ref{rings}. The first Monte Carlo snapshots of the contact zone show the formation of many small TCR/MHCp microclusters. At later times, the clusters close to the rim of the contact zone grow faster, and form an intermediate peripheral TCR/MHCp ring. The faster growth of the clusters close to the rim is caused by the diffusion of unbound TCR and MHCp molecules from the nonadhering membrane into the contact zone. Finally, the ring breaks to form a single large TCR/MHCp domain. 
 
{\it Regime 2}: For smaller TCR or MHCp concentrations, or smaller effective binding energy, we observe characteristic multifocal TCR/MHCp patterns at intermediate times. A typical example is shown in Fig.\ \ref{multifocal}. Initially, we observe again the nucleation of many TCR/MHCp microclusters throughout the contact zone. However, the overall area of TCR/MHCp domains now is not large enough for the formation of a TCR/MHCp ring. Instead, microclusters in the whole contact zone grow and coalesce, which leads to multifocal intermediates and finally again to a single TCR/MHCp domain.

\begin{figure}[b]
\hspace*{0.3cm}
\hspace*{-0.5cm}
\resizebox{0.9\columnwidth}{!}{\includegraphics{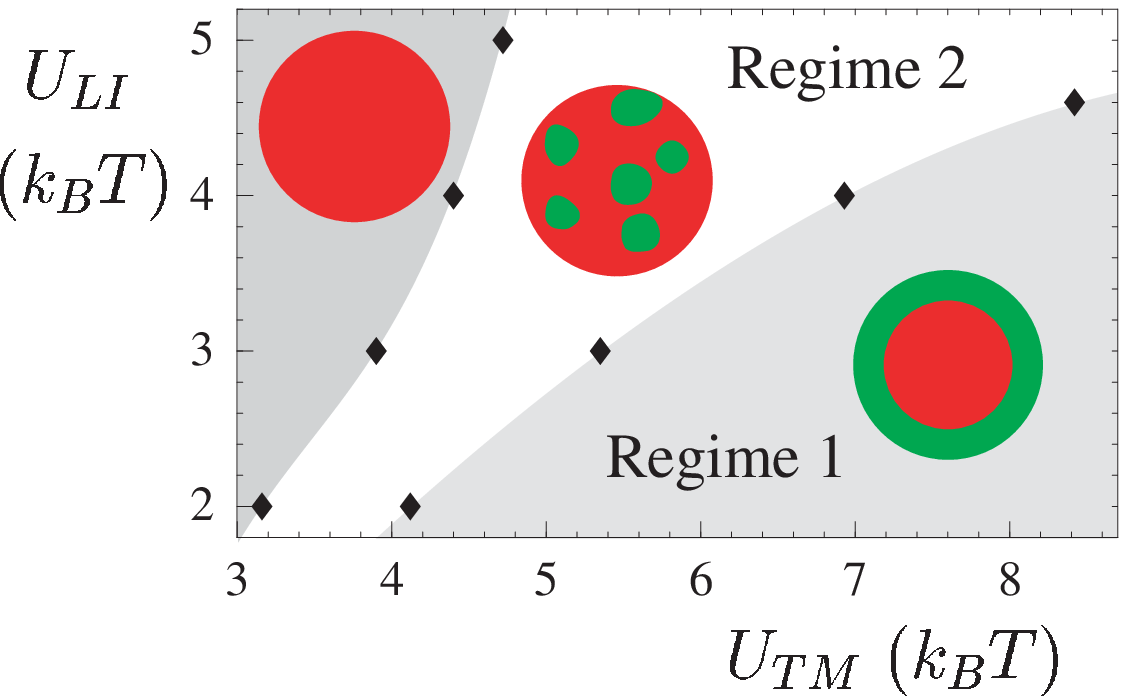}}

\vspace{0.3cm}

\resizebox{0.9\columnwidth}{!}{\includegraphics{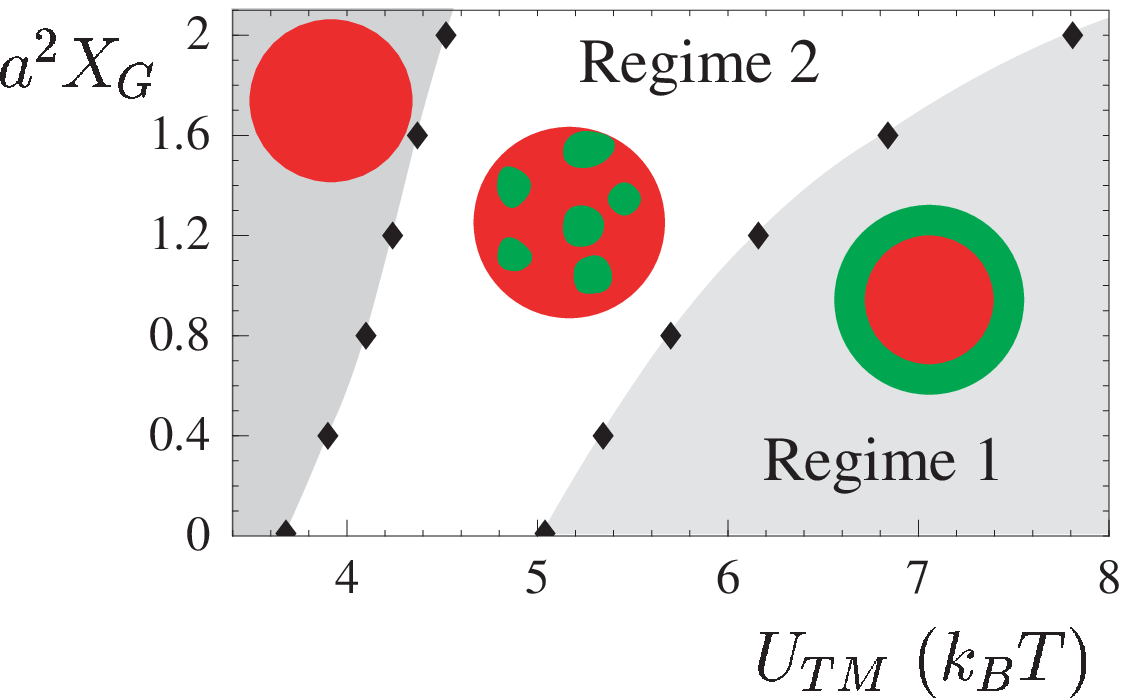}}
\caption{Dynamic regimes for T cell adhesion. The  concentrations of TCR, LFA-1, and ICAM-1 are $0.4 / a^2 \simeq 80$ molecules/$\mu$m$^2$ and the concentration of MHCp is $0.1 / a^2 \simeq 20$ molecules/$\mu$m$^2$. In the top diagram, the glycoprotein concentration in each of the membranes is $X_G=0.4/a^2$. In the bottom diagram, the binding energy $U_{LI}$ of LFA-1/ICAM-1 complexes has the value $3k_B T$. The black diamonds in the figure represent data points obtained from Monte Carlo simulations.  At large values of the binding energy $U_{TM}$ of the TCR/MHCp complexes, we observe a peripheral TCR/MHCp ring at intermediated times as in Fig.~\ref{rings} (Regime 1). At medium values of $U_{TM}$, multifocal patterns as in Fig.~\ref{multifocal} are obtained at intermediate times (Regime 2). At small values of $U_{TM}$, TCR/MHCp domains in the contact zone do not form. The threshold for the formation of TCR/MHCp domains and the crossover between the two dynamic regimes depend on the binding energy $U_{LI}$ of LFA-1/ICAM-1 complexes and the glycoprotein concentration $X_G$ in both membranes.\label{regimes}}
\end{figure}

To distinguish the two dynamic regimes systematically, we consider a peripheral ring of the contact zone with distances $r > 35a$ from the center, and divide this ring into 100 equal segments. For each Monte Carlo pattern obtained during adhesion, we determine the fraction $Y$ of ring segments which contain bound TCR/MHCp complexes. A fully closed peripheral TCR/MHCp ring corresponds to a ring occupation $Y=100$ \%. We find that the crossover between the two dynamic regimes can be appropriately described by a maximum ring occupation of $Y^\text{max}=80$~\% attained during adhesion. A pattern evolution with $Y^\text{max}<80$~\% typically has multifocal intermediates as in Fig.~\ref{multifocal} (Regime 2), while pattern evolutions with $Y^\text{max}>80$ exhibit the inverted synapse of T cells with  peripheral TCR/MHCp ring, see Fig.~\ref{rings} (Regime 1). 

The diagram at the top of Fig.~\ref{regimes} shows how the dynamic regimes for pattern formation depend on the effective binding energies $U_{TM}$ and $U_{LI}$ of the TCR/MHCp and LFA-1/ICAM-1 complexes. These binding energies are proportional to the logarithm of the `ideal' 2d equilibrium constants of the complexes, see Appendix. An increase in $U_{TM}$ in general leads to more TCR/MHCp complexes in the contact zone, while an increase in $U_{LI}$ leads to the binding of more LFA-1/ICAM-1 complexes. We observe three different scenarios: (i) At small values of $U_{TM}$, TCR/MHCp domains do not form at all in the contact zone which then is completely occupied by LFA-1/ICAM-1 complexes. TCR/MHCp domains only form above a threshold value for $U_{TM}$. This threshold value increases with $U_{LI}$. (ii) At large values of $U_{TM}$, we observe Regime 1 of pattern formation with the characteristic peripheral ring of TCR/MHCp complexes as in Fig.~\ref{rings}. (iii) At intermediate values $U_{TM}$, we find the patterns of Regime 2 with characteristic multifocal intermediates as in Fig.~\ref{multifocal}. The crossover value of $U_{TM}$ separating Regime 1 and Regime 2 increases with  $U_{LI}$. The intermediate TCR/MHCp ring of Regime 1 only forms if sufficiently large numbers of TCR/MHCp complexes are present in the contact zone. Instead of varying the effective binding energies $U_{TM}$ and $U_{LI}$, the numbers of bound receptor/ligand  complexes in the contact zone could also be changed by varying the overall concentrations of the receptors and ligands, with similar effects on the pattern formation.

The diagram at the bottom of Fig.~\ref{regimes} shows the effect of the glycoprotein concentration $X_G$ on the adhesion dynamics. The length of the glycoproteins is compatible with the length of the LFA-1/ICAM-1 complexes. Hence, the glycoproteins can enter the red LFA-1/ICAM-1 domains in the contact zone, but are excluded from the green TCR/MHCp domains. The accessible membrane area for the glycoproteins increases with the fraction of LFA-1/ICAM-1 domains in the contact zone, and so does the entropy of the glycoprotein distribution. Therefore, an increase in the overall glycoprotein concentrations  leads to a larger fraction of red LFA-1/ICAM-1 domains in the contact zone, and thus has a similar effect as increasing the binding energy $U_{LI}$ of the LFA-1/ICAM-1 complexes. 

\begin{figure}[t]
\resizebox{0.7\columnwidth}{!}{\includegraphics[angle=0]{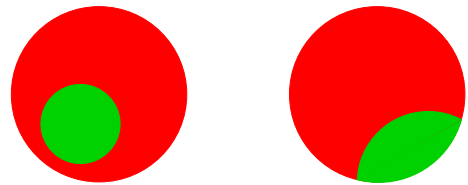}}
\caption{Two possible patterns with a single TCR/MHCp domain, shown in green. In the left pattern, the TCR/MHCp domain only has boundaries with the red LFA-1/ICAM-1 domain. In the right pattern, the TCR/MHCp domain is in contact with the rim of the contact zone. 
 }
\label{cartoonII}
\end{figure}

\begin{figure*}
\resizebox{2\columnwidth}{!}{\includegraphics{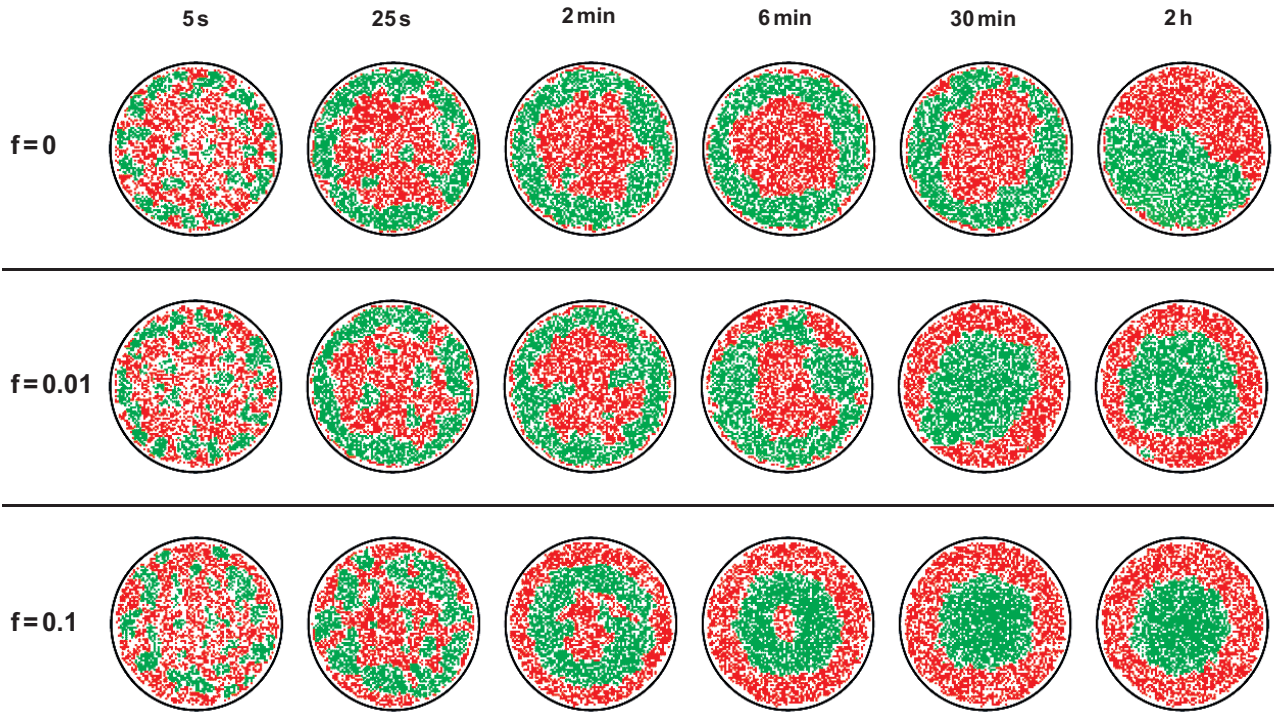}}
\vspace{-1cm}

\caption{Pattern formation with active transport of TCRs towards the center of the contact zone. Membrane patches with bound TCR/MHCp complexes are shown in green, patches with LFA-1/ICAM-1 complexes in red. Molecular concentrations and binding energies are the same as in Fig.~\ref{rings} (ring regime). (Top) At zero force, the intermediate TCR/MHCp pattern is stable for 30 minutes and more. In the final equilibrium pattern, both types of domains are in contact with the rim of the adhesion region, see section III. (Middle) At the force $F=0.01k_B T/a$, the final equilibrium state is the target-shaped mature synapse of T cells. This state is already established within 30 minutes. (Bottom) At the 10-fold stronger force  $F=0.1$ $k_B T/a$, the final target-shaped pattern already forms within 5 to 10 minutes. An intermediate pattern with a TCR/MHCp ring appears around 30 seconds after initial contact.
\label{patternsWithForce}}
\end{figure*}

In both dynamic regimes of pattern formation, the coalescence of clusters finally leads to a single TCR/MHCp domain in our model. In the absence of active transport processes, we always observe that the final TCR/MHCp domain is in contact with the rim of the contact zone,  see Figs.~\ref{rings} and \ref{multifocal}. This behavior can be understood from the line tensions at the domain boundaries and at the rim of the contact zone. The line tension between TCR/MHCp and LFA-1/ICAM-1 domains, $\lambda$, is an energy per unit length of the domain boundaries, and is mainly caused by the elastic energy of the membanes within the boundary regions between the domains. At the boundary, the membranes are bent to connect a TCR/MHCp domain with a membrane separation of around 15~nm with an LFA-1/ICAM-1 domain with a separation of around 40~nm. Similar line tensions, or energies per length, arise at the rim of the contact zone, both for TCR/MHCp and LFA-1/ICAM-1 domains adjacent to the rim, which we denote here by $\lambda_{TM}$ and $\lambda_{LI}$. Of special interest here is the difference $\lambda_r=\lambda_{TM}-\lambda_{LI}$, the energy per length for {\it replacing} an LFA-1/ICAM-1 boundary at the rim of the contact zone by a TCR/MHCp boundary. While $\lambda$ is always positive, reflecting the phase separation, $\lambda_r$ in principle can have both positive and negative sign. \footnote{These line tensions determine the contact angle $\theta$ of the green lense in the right pattern of Fig.~\ref{cartoonII} via $\cos(\theta)=(\lambda_{LI}-\lambda_{TM})/\lambda$.}

Let us assume that the final TCR/MHCp domain has a smaller area than the LFA-1/ICAM-1 domain, which seems to be the case for T cells. If $\lambda_r$ is larger than $\lambda$, boundaries of  TCR/MHCp domains inside the contact zone with the LFA-1/ICAM-1 domain are energetically more favorable than boundaries at the rim. Hence, the final TCR/MHCp domain should be circular and located anywhere inside the contact zone, to minimize the overall line tension, see Fig.~\ref{cartoonII}. In contrast, if $\lambda$ is larger than $\lambda_r$, TCR/MHCp domain boundaries at the rim of the contact zone are more favorable than interior boundaries, and the final TCR/MHCp domain should be in contact with the rim. In our simulations, $\lambda$ is clearly larger than $\lambda_r$, although $\lambda_r$ is positive since our boundary conditions, a rim separation of 100 nm, favor LFA-1/ICAM-1 domains at the egde of the contact zone. In the case of cells, it is reasonable to assume that $\lambda$ is much larger than  $\lambda_r$, since separation differences between 15 and 40 nm at the edge of the contact zone should not cause large energetic differences in the cell elasticity.

\section{Adhesion dynamics with active transport of TCRs}

In T cells, active processes transport receptors into the contact zone  \cite{wulfing98a} and glycoproteins out of this region \cite{allenspach01,delon01}. The framework enabling these transport processes is the actin cytoskeleton which polarizes during adhesion around the center of the contact zone \cite{alberts94,dustin98}. For TCRs, the transport is mediated by myosin, a molecular motor protein binding to the actin filaments. Here, we model the transport of TCRs as directed diffusion. For simplicity, we assume that each TCR molecule experiences a constant force which is directed towards the center of the contact zone midpoint. This force corresponds to an additional term $F\cdot r$ in the configurational energy of each TCR where $F$ is the magnitude of the force and $r$ the distance of the receptor from the center of the contact zone. Under the influence of this force, diffusive steps bringing TCRs closer to focal point of the cytoskeleton in the center of the contact are, in general, more likely than diffusive steps in the opposite direction.

\begin{figure}
\vspace*{-2cm}
\resizebox{\columnwidth}{!}{\includegraphics{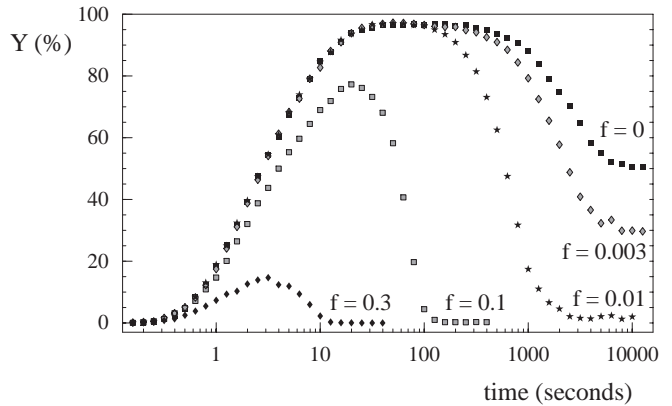}}
\caption{TCR/MHCp ring occupation $Y$ at various forces $F=f\cdot k_B T/a$ as a function of time $t$ for the same molecular concentrations as in Figs.\ \ref{rings} and \ref{patternsWithForce}. A ring occupation $Y=100$~\% corresponds to  fully closed peripheral ring of TCR/MHCp complexes, smaller percentages of $Y$ correspond to partial occupations of the peripheral ring with distances $r>35$ $a$ from the center of the contact zone. T-cell like pattern evolution with intermediate values of $Y\gtrsim 80$~\% (inverted synapse) and final values of $Y\simeq 0$~\% (`mature' synapse) are obtained for forces between $F=0.01 k_B T/a$ and $F=0.1 k_B T/a$, see also Fig.~\ref{patternsWithForce}. Larger forces prevent the formation of an intermediate peripheral TCR/MHCp ring, while smaller forces do not lead to a final central TCR/MHCp cluster. The data points represent averages over 24 independent Monte Carlo runs for each force.
\label{Yforce}}
\end{figure}

Figure \ref{patternsWithForce} compares the pattern evolution at zero force with patterns at the forces $F=0.01k_B T/a\simeq 6\cdot 10^{-16}$~N and $F=0.1k_B T/a\simeq 6\cdot 10^{-15}$~N. The concentrations and binding energies are the same as in Fig.~\ref{rings}. For these values, the force $F=0.01k_B T/a$ is close to the force threshold leading to a target-shaped final synapse with central TCR/MHCp cluster, see Fig.~\ref{Yforce}. Besides leading to a central TCR/MHCp cluster, the active forces speed up the pattern evolution. At the weaker force $F=0.01k_B T/a$, the final equilibrium state is reached after approximately 30 minutes, while the 10-fold stronger force $F=0.1k_B T/a$ leads to equilibrium within few minutes. The absolute times are based on the estimate that one Monte Carlo step roughly corresponds to 1 ms, see above.
 However, TCR/MHCp rings at intermediate times form in all three cases shown in Fig.~\ref{patternsWithForce}. Only significantly stronger forces prohibit the formation of intermediate rings by pulling the TCRs more directly to the center of the contact zone. 

To quantify the impact of the active forces on the pattern evolution, we consider again the peripheral ring occupation $Y$ of the TCR/MHCp complexes, see Fig.~\ref{Yforce}. Values $Y\gtrsim 80$~\% at intermediate times indicate a peripheral TCR/MHCp ring as in the inverted T cell synapse, and values  $Y\simeq 0$~\% at later times correspond to target-shaped patterns with a central TCR/MHCp domain as in the `mature' T cell synapse. We consider the same molecular concentrations and binding energies as in the Figs.~\ref{rings} and \ref{patternsWithForce}. For these values, active forces $F\lesssim 0.003k_B T/a$ are too weak to cause a final central TCR/MHCp cluster, while forces $F > 0.1k_B T/a$ guide the TCRs very quickly to the contact zone center, which prevents the formation of a peripheral TCR/MHCp ring at intermediate times. We obtain a T-cell like pattern evolution for forces $0.01k_B T/a \lesssim F \lesssim 0.1 k_B T/a$ with an intermediate inverted synapse and a final `mature' synapse exhibiting a central TCR/MHCp cluster. Experimentally, the mature synapse of T cells has been observed to form on timescales  between 5 and 30 minutes, which agrees with the equilibration times we obtain for these forces. In the absence of active forces ($F=0$), the intermediate peripheral TCR/MHCp ring seems to be metastable and appears in our simulations for times up to an hour. This metastability might explain the inverted NK cell synapse  which consists of a peripheral ring of short receptor/ligand complexes, and a central domain containing the longer integrins. The inverted synapse of NK cells seems to formed by self-assembly, since it is not affected by ATP depletion or cytoskeletal inhibitors \cite{davis99,fassett01}.

\section{Conclusions} 

In this article, we have considered  possible mechanisms underlying the pattern formation during T cell adhesion. We propose a novel mechanism for the formation of {\em intermediate} patterns, which is based on the nucleation of TCR/MHCp microdomains throughout the contact zone and the diffusion of free receptors and ligands into this zone. This self-assembly mechanism does not require active processes and leads to the inverted synapse pattern of T cells if the TCR/MHCp concentration is large enough. The mechanism leads to multifocal intermediates as observed during thymozyte adhesion for smaller TCR/MHCp concentrations. 

According to our model, the {\em final} T cell pattern with a central TCR/MHPp domain is caused by active transport of TCRs towards the center of the contact zone, where the polarized cytoskeleton has a focal point. This seems to be in agreement with experiments which show that the central TCR/MHCp movement is inhibited by blocking the active cytoskeletal transport with Cytochalasin D \cite{grakoui99}.
 
In our model, the coalescense of  domains eventually leads to a single TCR/MHCp domain. Without active transport, we find that this domain is located at the rim of the contact zone. We obtain the final bull-eye pattern of the mature synapse with a central TCR/MHCp domain only in the presence of active TCR transport. For simplicity, we have modeled the active transport by a constant average force F which acts on the TCRs and is directed towards to the center of the contact zone.  We obtain the characteristic pattern inversion of T cells for average forces in the range $6\cdot 10^{-16}$~N $\lesssim F \lesssim 6\cdot 10^{-15}$~N. Smaller average forces do not lead to the mature T cell synapse with central TCR/MHCp cluster, while larger forces disrupt  the peripheral TCR/MHCp ring of the intermediate, inverted synapse. It is important to note that these {\em average} forces acting on a single TCR are significantly smaller than the maximum forces around 1~pN$=10^{-12}$~N which can be exerted by a single molecular motor. The transport of a TCR molecule over larger distances presumably involves several cytoskeletal binding and unbinding events. Since the TCRs are membrane-anchored, the transport along cytoskeletal fibers close to the membrane may also be affected by additional friction within the membrane.

Active cytoskeletal processes could also stabilize the multifocal patterns of thymozytes which have been observed by Hailman et al. The cytoskeleton of thymozytes presumably remains in a mobile, nonpolarized state which still allows cell migration \cite{hailman02}. The few TCR/MHCp clusters of thymozytes may be coupled to the cytoskeleton, thus following its movements. An alternative explanation for the multifocal patterns has been given in terms of critical fluctuations \cite{sjlee03,raychaudhuri03}. Fluctuations close to a critical point can lead to the appearance and disappearance of small domains, since the line tension of the domain boundaries then is  low. However, the few small TCR/MHCp clusters observed bei Hailman et al.\ are rather circular, which indicates a relatively large line tension. In addition, the multifocal patterns were observed over a 100-fold range of antigen concentrations \cite{hailman02}, whereas critical fluctuations can only be observed in a rather narrow concentration range close to the critical point. 

Natural killer (NK) cells form an inverted synapse, consisting of a peripheral ring of short HLA-C/KIR complexes and a central domain with the longer LFA-1/ICAM-1 complexes. The formation of the NK cell synapse seems not to depend on active cytoskeletal processes, since ATP depletion or disruption of the cytoskeleton has no effect on the pattern \cite{davis99,fassett01}. A possible explanation for the NK cell synapse is the metastability of the inverted pattern in the absence of active cytoskeletal processes. Without active transport, the inverted intermediate synapse persists up to an hour in our model. As mentioned above, experiments show that the central TCR/MHCp movement of T cells requires cytoskeletal transport as well. 

We have applied our model here to T cell adhesion, using the specific lengths of the TCR/MHCp and ICAM-1/LFA-1 complexes in the interaction potentials (\ref{potTM}) and (\ref{potLI}). However, the model is rather general and also applies to other cell adhesion events. We have previously considered a simpler membrane system with `stickers' and `repellers' \cite{weikl02a}. The phase separation into sticker- and repeller-rich domains is driven by the length difference between the two molecule types. In the cell adhesion geometry, we obtained intermediate patterns which are similar to those presented here. A difference to T cell membranes is that the repeller-rich domains are unbound. Large-scale membrane fluctuations in these domains then drive the final sticker clusters towards the center of the contact zone, at least for the `free' boundary conditions with unconstrained membrane separation at the contact zone rim  \cite{weikl02a}. In contrast, the coexisting TCR/MHCp and ICAM-1/LFA-1 domain types of T cells are both bound, and large-scale membrane fluctuations are suppressed. 

We have characterized the receptor/ligands by an effective binding energy which is directly related to an effective 2D equilibrium constant $K_{2D}$, see Appendix. The equilibrium constant $K_{2D}$ is the ratio of the kinetic on- and off-rates $k_{on}$ and $k_{off}$ for the receptor-ligand binding \cite{bell78}. Characterising the binding kinetics by the single parameter $K_{2D}$ rather than the two parameters $k_{on}$ and $k_{off}$ is justified at least if the on-reaction of a receptor-ligand pair in apposing membrane patches of our discrete model is faster than the timescale 1~ms for the diffusive Monte Carlo steps. In general, such an approach may also be justified by a local equilibration within domains. 
 
\begin{appendix}   
\section{Binding energies and 2d equilibrium constants}

In this appendix, we consider the relation between the 2d equilibrium constants and the binding energies of receptors and ligands with square-well potentials (\ref{potTM}) or  (\ref{potLI}). The 2d equilibrium constants are defined by   
\begin{equation}
K_{2d} = \frac{x_{RL}}{ x_R x_L } \label{K2d}
\end{equation}
where $x_R$ is the area concentration of receptors R in one of the membranes, $x_L$ is the area concentration of the ligands L in the apposing membrane, and $x_{RL}$ is the concentration of the complexes. In general, $K_{2d}$ depends on the state of the membrane, not only on the interaction potential of receptors and ligands. For example, if the membrane separation is to large to allow complex formation, the equilibrium constant (\ref{K2d}) is zero. In a bound state, $K_{2d}$ will depend on the separation and roughness of the membrane, which in turn are affected by the concentrations of the receptors, ligands, and steric repellers such as glycoproteins. Here, we only consider the `ideal' 2d equilibrium constant for a membrane segment which is entirely within the binding range of the receptor/ligand square-well interaction. 

For convenience, we choose in the following the grandcanonical ensemble with chemical potentials $\mu_R$ and $\mu_L$ for receptors and ligands. However, the ideal 2d equilibrium constant derived below will be independent of $\mu_R$ and $\mu_L$ and thus applies also to the canonical ensemble with fixed overall receptor and ligand concentrations. 

Let us consider two apposing membrane patches within binding range of the receptors R and ligands L.  A `state' of these apposing patches  is characterized by the numbers $m_L$  and $m_R$ of ligands and receptors present in the two membranes. The configurational energy is $h(m_L,m_R)=U_{RL} \min(m_L,m_R)-\mu_L m_L -\mu_R m_R$ where $U_{RL}$ is the binding energy of RL complexes. To simplify the notation below, the parameters $\mu_R$,  $\mu_L$, and $U_{RL}$ are taken to be in units of the thermal energy $k_B T$. In a given state, there can be $k$ bound RL complexes, plus either  $i$ uncomplexed ligands or $j$ uncomplexed receptors (or no additional uncomplexed molecules). The partition function then has the form
\begin{eqnarray}
z &=& \sum_{m_L=0}^\infty \sum_{m_R=0}^\infty \exp(-h(m_L,m_R))\nonumber\\
&=&\sum_{k=0}^\infty e^{k(\mu_L+\mu_R-U_{RL})}
\bigg(1+\sum_{i=1}^\infty e^{i\, \mu_L} + \sum_{j=1}^\infty e^{j\, \mu_R}\bigg)\nonumber\\
&=& \frac{1}{1-e^{\mu_L + \mu_R - U_{RL}}}
\bigg(\frac{1}{1-e^{\mu_L}} + \frac{1}{1-e^{\mu_R}} - 1\bigg) 
\end{eqnarray}
and the concentration of bound RL complexes is
\begin{eqnarray}
x_{RL} = \frac{1}{a^2z}\sum_{k=1}^\infty k e^{k(\mu_L+\mu_R-U_{RL})}
\bigg(1+\sum_{i=1}^\infty e^{i\, \mu_L} 
\nonumber\\
+ \sum_{j=1}^\infty e^{j\, \mu_R}\bigg)
=\frac{e^{\mu_L+\mu_R-U_{RL}}}{a^2(1-e^{\mu_L+\mu_R-U_{RL}})}
\end{eqnarray} 
In deriving these expressions we made use of
\begin{equation}
\sum_{k=0}^\infty s^k =\frac{1}{1-s} \;\; , \;\;\;\;\; \sum_{k=0}^\infty k s^k = \frac{s}{(1-s)^2}
\end{equation}
for $s<1$.
The concentration of uncomplexed receptors is given by 
\begin{eqnarray}
x_R &=& \frac{1}{a^2 z} \sum_{k=0}^\infty e^{k(\mu_L+\mu_R-U_{RL})}
\sum_{j=1}^\infty j e^{j\, \mu_R}\nonumber\\
&=& \frac{1}{a^2}\left(\frac{1}{1-e^{\mu_R}} - \frac{1}{1-e^{\mu_R+\mu_L}}\right)
\end{eqnarray}
and accordingly, the concentration of uncomplexed ligands is
\begin{eqnarray}
x_L = \frac{1}{a^2}\left(\frac{1}{1-e^{\mu_L}} - \frac{1}{1-e^{\mu_R+\mu_L}}\right)
\end{eqnarray}
Hence, the ideal 2-dimensional equilibrium constant of receptors and ligands in apposing membrane patches within binding range is given by  
\begin{equation}
\fbox{$\displaystyle
K_{2d} = \frac{x_{RL}}{ x_L x_R } = 
\frac{a^2\left(1-e^{\mu_L + \mu_R}\right)^2}
   {e^{U_{RL}} - e^{\mu_L + \mu_R}} \simeq a^2 e^{-U_{RL}}$}
\end{equation}  
The last expression holds for $e^{\mu_L + \mu_R}\ll 1$ which is true for the receptor and ligand concentrations studied in this article, see below. 
   
For comparison, let us also consider two apposing membrane patches with a separation outside of the binding range of receptors and ligands.  The partition function for the patches then is
\begin{equation}
z = \sum_{i=0}^\infty  e^{i\, \mu_L }\sum_{j=0}^\infty  e^{j\, \mu_R}
=\frac{1}{(1-e^{\mu_L})(1-e^{\mu_R})} \hspace*{0.2cm}
 \end{equation}
and the concentrations of receptors and ligands are given by
\begin{equation}
x_R = \frac{\sum_{j=1}^\infty j e^{j\, \mu_R}}{a^2\sum_{j=0}^\infty  e^{j\, \mu_R}}
= \frac{e^{\mu_R}}{a^2\left(1-e^{\mu_R}\right)}
\; , \;\; x_L= \frac{e^{\mu_L}}{a^2\left(1-e^{\mu_L}\right)}
\end{equation}

\end{appendix}

\end{document}